\documentclass[varenna]{cimento}
\usepackage{graphicx} 
\usepackage{amssymb}

\title{A new geometrical approach to void statistics}
\author{M.~C. Werner}
\institute{Kavli Institute for the Physics and Mathematics of Universe (WPI) \\  University of Tokyo \\
5-1-5 Kashiwanoha, Kashiwa 277-8583, Japan \\ Department of Mathematics \\ Duke University \\ Science Drive, Durham NC 27708, USA}
\begin{document}

\maketitle

\begin{abstract}
Modelling cosmic voids as spheres in Euclidean space, the notion of a de-Sitter configuration space is introduced. It is shown that a uniform distribution over this configuration space yields a power-law approximating the void size distribution in an intermediate range of volumes, as well as an estimate for the fractal dimension of the large scale structure.
\end{abstract}

\section{Introduction}
Cosmic voids play an increasingly important r\^{o}le in cosmology (see, e.g., \cite{LaWa10}). A new geometrical framework to study the distribution of voids proposed in ref.~\cite{GiWeYoCh} was presented at the Enrico Fermi School \textit{New Horizons for Observational Cosmology} in Varenna. This approach models voids as spheres in 3-dimensional Euclidean space, which are represented by points in a 4-dimensional configuration space with de-Sitter geometry. It turns out that a uniform distribution over this configuration space gives a good approximation to the void size distribution in an intermediate volume range, and it can also be used to estimate the fractal dimension of the complementary set of the voids, that is, the large scale structure.
\newline
Hence, this is a new application of de-Sitter geometry to cosmology which derives, however, from classical sphere geometry rather than general relativity. Another application of this de-Sitter configuration space of spheres to astronomy in a different context is discussed in ref.~\cite{GiWe13}. 

\section{De-Sitter configuration space}
\subsection{From sphere geometry to de-Sitter}
\begin{figure}
\centering
\begin{minipage}{0.5\textwidth}
  \centering
  \vspace{5mm}
  \includegraphics[width=0.9\columnwidth]{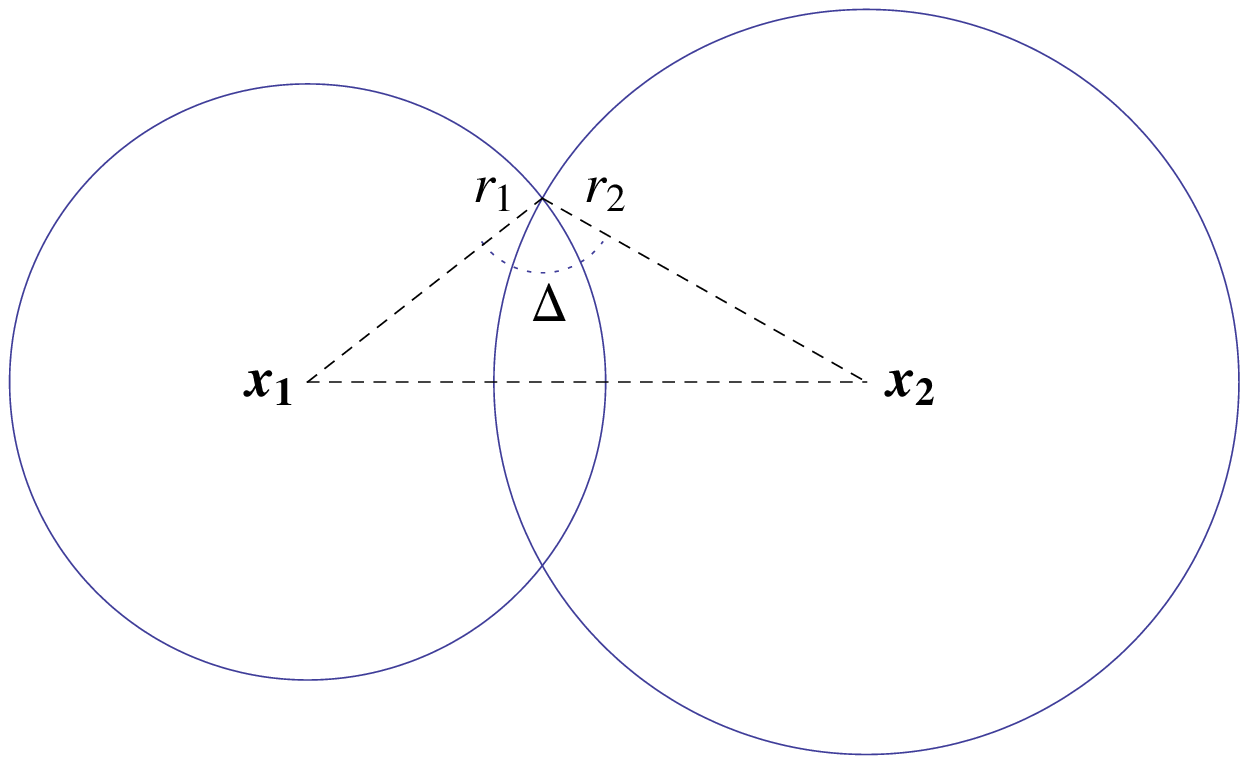}
  \vspace{5mm}
  \caption{Intersection angle of spheres.}
\label{fig:spheres}
\end{minipage}%
\begin{minipage}{0.5\textwidth}
  \centering
  \rotatebox{-90}{
  \includegraphics[width=0.7\columnwidth]{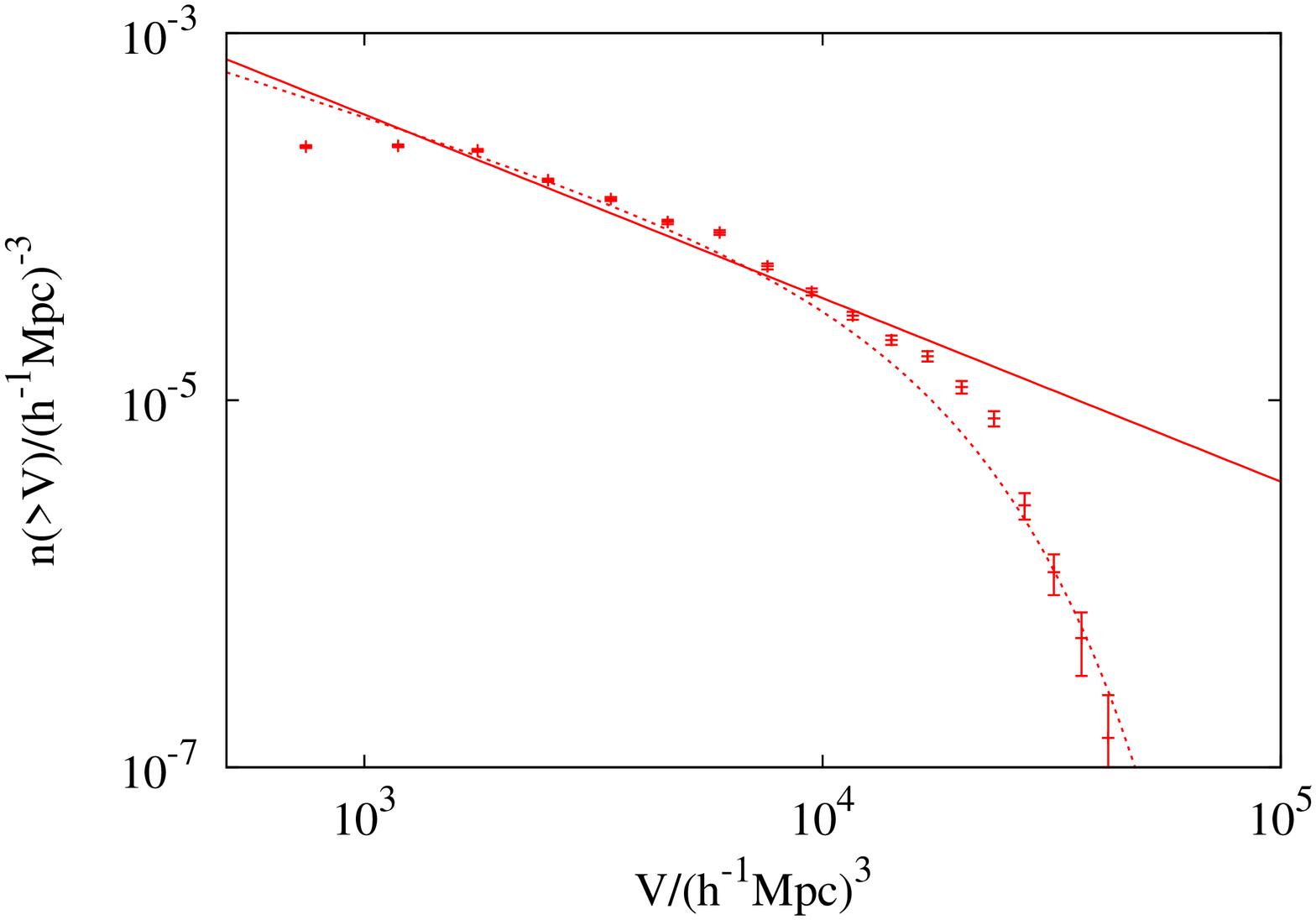}}
  \caption{Cumulative number density of voids.}
\label{fig:voids}
\end{minipage}
\end{figure}

In the 3-dimensional Euclidean space $\mathbb{E}^3$ endowed with the usual Euclidean metric $\delta_{ij}$, an unoriented sphere can be uniquely defind by its centre at $\mathbf{x}=(x^i) \in \mathbb{E}^3,\ 1\leq i\leq 3,$ and its radius $r>0$, so that the tuple $y^\mu=(r,x^i),\ 0\leq \mu\leq 3,$ defines its configuration and can be used as a coordinate in a 4-dimensional configuration space. We would like to introduce a notion of distance on this configuration space such that spheres which overlap in the same way have the same distance on the configuration space, and this overlap can be characterized by the angle of sphere intersection $\Delta$. Given two spheres $(r_1,x^i_1)$ and $(r_2,x^i_2)$, this is defined as
\begin{equation}
\cos \Delta=\frac{r_1^2+r_2^2-\delta_{ij}(x_1^i-x_2^i)(x_1^j-x_2^j)}{2r_1r_2},
\label{delta}
\end{equation}
and is illustrated in fig.~\ref{fig:spheres}. In other words, we seek a metric on the configuration space whose isometries preserve $\Delta$, and find that this is provided by de-Sitter geometry. To see this, a short excursion to the sphere geometry of Lie and Laguerre, which studies contact relations between spheres, is in order. Consider the set of Lie cycles which consists of all oriented spheres, called Laguerre cycles, all oriented hyperplanes, and the object infinity. It turns out (e.g., \cite{Be12}, proposition 3.56) that one can assign bijectively to every Lie cycle a homogeneous Lie cycle coordinate
\[
\left[\lambda^0,\lambda^1,\lambda^2,\lambda^3,\lambda^4,\lambda^5\right]=\left[k\lambda^0,k\lambda^1,k\lambda^2,k\lambda^3,k\lambda^4,k\lambda^5\right], \quad \mbox{any} \quad k\neq0,  
\]
which obeys the Lie quadric,
\begin{equation}
-(\lambda^0)^2+(\lambda^1)^2+(\lambda^2)^2+(\lambda^3)^2+(\lambda^4)^2-(\lambda^5)^2=0.
\label{quadric}
\end{equation}
For a Laguerre cycle given by $(r,x^i), \ r\neq 0$, the corresponding Lie cycle coordinate may be written (cf. \cite{Be12}, p. 154)
\begin{equation}
\left[\frac{-r^2+\delta_{ij}x^ix^j+1}{2},x^1,x^2,x^3,\frac{-r^2+\delta_{ij}x^ix^j-1}{2},-r\right]:=\left[X^0,X^1,X^2,X^3,X^4,1\right].
\label{lie}
\end{equation}
Now suppose that the $X^A,\ 0\leq A \leq 4$, are components of a coordinate in 5-dimensional Minkowski space $\mathbb{E}^{1,4}$, with the standard metric $\eta_{AB}=\mathrm{diag}(-1,1,1,1,1)$, then the Lie quadric (\ref{quadric}) implies that this denotes a point on a de-Sitter quadric hypersurface in this Minkowski space,
\[
dS^4: \quad \eta_{AB}X^AX^B=1, \quad X=(X^A) \in\mathbb{E}^{1,4},  
\]
which defines a 4-dimensional de-Sitter space in the usual way. Returning to our original problem of unoriented spheres in 3-dimensional Euclidean space, then, we see that each can be represented by precisely one point in (half of the full) de-Sitter space. Moreover, given two spheres defined by $X^A_1, \ X^B_2$, a little algebra shows that, in fact,
\[
\eta_{AB}X^A_1 X^B_2|_{dS^4}=\cos \Delta, 
\]
so that this space has the desired metric property as well. Hence, we can identify 4-dimensional de-Sitter space as a suitable configuration space for spheres, as promised. The metric $g$ on the de-Sitter quadric induced by the ambient Minkowski metric can be found from the line element
\[
\drm s^2=\eta_{AB} \drm X^A \drm X^B|_{dS^4}=\frac{1}{r^2}\left(-\drm r^2+\delta_{ij}\drm x^i\drm x^j\right):= g_{\mu\nu}\drm y^\mu\drm y^\nu,
\]
using a chart with configuration space coordinates $y^\mu=(r,x^i)$, so that the configuration space metric in the coordinate-induced basis is given by 
\begin{equation}
g_{\mu\nu}=\mathrm{diag}\left(-\frac{1}{r^2},\frac{1}{r^2},\frac{1}{r^2},\frac{1}{r^2}\right).
\label{metric}
\end{equation}
Hence, from this point of view, the sphere radius is a timelike coordinate and the sphere centre position a spacelike coordinate.

\subsection{From the uniform distribution to fractals}
Let us now turn to the uniform distribution of spheres over this de-Sitter configuration space, such that the number $\drm N$ of spheres with radius in the interval $[r,r+\drm r]$ and centre in the volume element $\drm v=\drm x^1\drm x^2 \drm x^3$ of Euclidean 3-space is proportional to the volume element of the configuration space $\drm v_g$, which can be computed from eq.~(\ref{metric}). Hence,
\[
\drm N \propto \drm v_g=\sqrt{-\det g}\drm y^0 \drm y^1 \drm y^2 \drm y^3=\frac{1}{r^4}\drm r\drm v.
\]
One can recast this in terms of the sphere volume $V=4\pi r^3/3$ to find the differential number density $\nu(V)$ in Euclidean 3-space of spheres with volumes in $[V,V+\drm V]$, as well as the corresponding cumulative number density $n(>V)$,
\begin{eqnletter}
\nu(V)& = & \frac{\drm N}{\drm V\drm v}\propto \frac{1}{V^2}, \label{diff} \\
n(>V) & = & \int_V^\infty \nu(V') \drm V':=\frac{C}{V}, \label{cumul}
\end{eqnletter}
respectively, where $C>0$ is a dimensionless constant. Our uniform distribution allows any sphere position and, therefore, any overlap. Hence, it is instructive to compare it to Mandelbrot's random spherical cutout model, which provides a geometrical interpretation of the constant $C$ in eq.~(\ref{cumul}). Consider a homogeneously filled cube of volume $V_0$ in $\mathbb{E}^3$, in which a sequence of spherical regions of volume $V_i$ are emptied, whose centres are placed randomly (distributed uniformly and independently) in the cube with faces identified. Now if, in an infinite sequence, the sum of volumes (regardless of overlap) diverges only slowly, namely as a harmonic series,
\begin{equation}
V_i=\frac{\tilde{C}V_0}{i}, \quad \mbox{so that} \quad \sum_{i=1}^k V_i\rightarrow \tilde{C}V_0\ln k \quad \mbox{as} \quad k\rightarrow \infty, 
\label{fractal1}
\end{equation}
then the cube will not be emptied entirely and there remains, in general, a complementary set of fractal dimension $D$ (cf. \cite{Fa97}, proposition 8.8, for the technical argument in $\mathbb{E}^1$). We can estimate $D$ by applying a box-counting algorithm, assuming for simplicity that, at the $k$th stage, $V_0$ is divided into $k$ equal boxes of edge length proportional to $k^{-\frac{1}{3}}$, and that overlap is still negligible. Then the remaining volume of the cube is $V(k)\simeq V_0(1-\tilde{C}\ln k), \ \tilde{C}\ln k\ll 1,$ occupying $N(k)\simeq k(1-\tilde{C}\ln k)$ boxes, and so the fractal dimension is
\begin{equation}
D=\frac{\log N(k)}{\log k^\frac{1}{3}}\simeq 3+3\frac{\ln (1-\tilde{C}\ln i)}{\ln i}\simeq 3(1-\tilde{C}).
\label{fractal2}
\end{equation}
Hence, for a volume $V\ll V_0$ whose corresponding $k \gg 1$, the cumulative number density of the spherical cutouts (again regardless of overlap) can now be written 
\begin{equation}
n(>V)=\frac{k-1}{V_0}\simeq \frac{\tilde{C}}{V}\simeq \left(1-\frac{D}{3}\right)\frac{1}{V}, 
\label{cumul2}
\end{equation}
using eq.~(\ref{fractal1}) and eq.~(\ref{fractal2}), which turns out to hold more generally (cf. \cite{Ma82}, p. 302). Comparing the cumulative number densities of eq.~(\ref{cumul}) and eq.~(\ref{cumul2}), we see that the uniform distribution over the de-Sitter configuration space results in the same scaling behaviour as the random spherical cutout model, both of which allow any sphere overlap. Thus identifying $\tilde{C}=C$, we obtain a geometrical interpretation of the proportionality constant
\begin{equation}
 C= 1-\frac{D}{3}
\label{c}
\end{equation}
in eq.~(\ref{cumul}) in terms of the fractal dimension $D$ of the complementary set not contained in the spheres. We shall now discuss the applicability of this geometrical approach to cosmic voids.

\section{Application to cosmology}
\subsection{Comparison with structure formation theory}
The excursion set formalism pioneered by Press and Schechter \cite{PrSc74} considers small Gaussian fluctuations of an almost homogeneous density field at a sufficiently early time, in which a window of length scale $L$ around an overdensity will gravitationally collapse to form an object of mass $M\propto L^3$ if the density contrast exceeds some critical value, $\delta(L)>\delta_{\mathrm{crit}}$. The fluctuation variance may depend on $L$ as well, $\sigma^2\propto L^{-k}, \ k \geq 0$, resulting in the differential number density
\begin{equation}
\nu(M)\propto \left(\frac{M}{M_0}\right)^{\frac{k}{6}-2}\exp\left[-\left(\frac{M}{M_0}\right)^\frac{k}{3}\right], 
\label{pressschechter}
\end{equation}
where $M_0$ is the mass scale of the exponential cutoff. As pointed out by Sheth and van de Weygaert \cite{ShvdWe04}, due to the symmetry of Gaussian fluctuations about the average density, the same reasoning applies also to underdensities and hence to voids of volume $V\propto L^3$. Therefore, given a scale-free variance ($k=0$) and void volumes sufficiently smaller than the exponential cutoff, we recover the same differential number density of voids from structure formation theory,
\[
\nu(V)\propto V^{-2},
\]
as from the uniform de-Sitter distribution, eq. (\ref{diff}). Of course, this ignores effects of void hierarchy, that is merging of voids in the cosmological evolution (cf. again ref.~\cite{ShvdWe04}), which cuts the power-law off at small volumes. Hence, the uniform de-Sitter distribution should hold approximately for cosmic voids of spherical shape, which is also supported theoretically (e.g., \cite{Ic84}), in an intermediate range of volumes.

\subsection{Comparison with data}
To illustrate this with data, an N-body simulation of $256^3$ particles in a box of edge length $240h^{-1}$Mpc with periodic boundary conditions was performed, using the GADGET-2 code \cite{Sp05} and a cosmology consistent with WMAP-7. After applying adaptive smoothing, spherical voids were identified and no merging algorithm was applied (cf. \cite{Co-etal05}) in order to allow any overlap as discussed above. The resulting cumulative number density of voids is shown in fig.~\ref{fig:voids}, together with an analytical fit (dotted curve) adapted from ref.~\cite{vBeMu08} and the power-law (solid line) from the uniform de-Sitter distribution, eq. (\ref{cumul}) with $C=0.36$. Given the approximate agreement with the data in an intermediate range of volumes, as expected, we can now also compute the fractal dimension of the large scale structure regarded as the complementary set of the voids, to obtain $D\simeq 1.9$ from eq. (\ref{c}) as a reasonable estimate (c.f. \cite{Jo-etal04}). Notice that the present approach differs from other fractal models using disjoint cutouts (e.g., \cite{Ga06}). 

\section{Outlook}
The next step in extending this geometrical idea is to investigate non-uniform distributions over the de-Sitter configuration space, which will result in deviations from the power-law (\ref{diff}). As mentioned above, it is those outside of the intermediate power-law range which will be most interesting for cosmology: the exponential cutoff at large volumes, since the largest voids may be particularly sensitive to cosmological parameters; and effects of void hierarchy and merging at small volumes. 

\acknowledgments
This work was supported by the World Premier International Research Center Initiative (WPI Initiative), MEXT, Japan.

\end{document}